\documentclass[letter]{aa}
\usepackage{txfonts}
\usepackage{graphicx}
\usepackage{natbib}
\bibpunct{(}{)}{;}{a}{}{,}
\begin{document}

\title{Hinode observations reveal boundary layers of magnetic elements 
       in the solar photosphere}
\author{R.~Rezaei\inst{1}, O.~Steiner\inst{1},  S.~Wedemeyer-B\"ohm\inst{2}, 
            R.~Schlichenmaier\inst{1}, W.~Schmidt\inst{1}, and  B.W.~Lites\inst{3}}
\institute{Kiepenheuer-Institut f\"ur Sonnenphysik, 
Sch\"oneckstrasse 6, D-79104 Freiburg, Germany 
\and
Institute of Theoretical Astrophysics, P.O. Box 1029, Blindern,
N-0316 Oslo, Norway
\and
High Altitude Observatory, NCAR, P.O. Box 3000, Boulder, CO 80307, USA\\
E-mail: [rrezaei; steiner; schliche; wolfgang]@kis.uni-freiburg.de, 
        sven.wedermeyer-bohm@astro.uio.no, lites@ucar.edu}
\date{Received 27 July 2007 / Accepted 31 October 2007}

\titlerunning{Variation of area asymmetry across a magnetic element}
\authorrunning{Rezaei et al. }

\abstract
{}
{We study the structure of the magnetic elements in network-cell interiors.}
{A quiet Sun area close to the disc centre was observed with the spectro-polarimeter
of the Solar Optical Telescope on board the Hinode space mission, which yielded 
the best spatial resolution ever achieved in polarimetric data of the 
\ion{Fe}{i}\,630\,nm line pair. For comparison and interpretation, we synthesize 
a similar data set from a three-dimensional magneto-hydrodynamic simulation.}
{We find several examples of magnetic elements, either roundish (tube) 
or elongated (sheet), which show a central area of negative Stokes-$V$ area asymmetry 
framed or surrounded by a peripheral area with larger positive asymmetry. 
This pattern was predicted some eight years ago on the basis of numerical simulations.
Here, we observationally confirm its existence for the first time.}
{We gather convincing evidence that this pattern of Stokes-$V$ area asymmetry 
is caused by the funnel-shaped boundary of magnetic elements that separates the flux 
concentration from the weak-field environment. On this basis, we conclude that 
electric current sheets induced by such magnetic boundary layers are common in the 
photosphere.}


\maketitle

%%%%%%%%%%%%%%%%%%%%%%%%%%%%% Introduction %%%%%%%%%%%%%%%%%%%%%%%%%%%%%%%

\section{Introduction}

With ever increasing polarimetric sensitivity and spatial resolution, it becomes
now clear that even the most quiescent areas on the solar surface harbour ample
amounts of magnetic flux. This flux becomes visible in small patches of
field concentrations, called magnetic elements for short. While horizontal 
fields tend to occur near the edge of granules, the present investigation 
focusses on fields predominantly oriented in the vertical direction and mainly
occurring in the intergranular space \citep{lites_etal_07b,lites_etal_07a}. There, 
magnetic elements are often visible in G-band filtergrams as point-like objects.
In more active regions, magnetic elements of extended size appear
in the form of `crinkles', `ribbon bands', `flowers', etc. \citep{berger+al04}, 
which show a sub-structure consisting of a dark central area surrounded by a
striation of bright material and further outside by a 
downdraft of plasma \citep{langangen_etal_07}. While Doppler measurements
of this downdraft are at the limit of spatial resolution with
the best ground-based solar telescopes, spectro-polarimetric 
measurements have less spatial resolution due to image degradation by
atmospheric seeing over the required long integration times. 

Measurements with the spectro-polarimeter on board the Hinode satellite
have a superior spatial resolution of  approximately $0.3\arcsec$ thanks to 
excellent pointing capabilities \citep{shimizu_etal_07} and the absence of 
seeing. Using Hinode data, we show examples of unipolar magnetic 
elements of the network-cell interiors that possess a distinct 
sub-structure, which is strikingly manifested in maps of the asymmetry of 
spectral lines in the circularly polarized light --- the Stokes-$V$ area 
asymmetry. 

The areas of the blue and the red lobes of Stokes-$V$ profiles are identical
when formed in a static atmosphere, but become asymmetric
in the presence of gradients in the velocity and magnetic field strength
\citep{illing+al75,solanki+pahlke88,sanchez-almeida+al89,landolfi+landi96}. 
Hence, the variation in the Stokes-$V$ asymmetry across magnetic elements 
gives information on their magnetic field and plasma flow properties.
Since this information is involved, we computed synthetic Stokes-$V$
profiles of magnetic elements that occur in a three-dimensional simulation
of magneto-convection and compare their asymmetry with the measured ones
in order to understand their origin.

%%%%%%%%%%%%%%%%%%%%%%%% Observations & Data Reduction %%%%%%%%%%%%%%%%%%%%%%%%%%

\section{Observation and data reduction}

The observations were carried out on 10 March, 2007 with the
spectro-polarimeter (SP) of the Solar Optical Telescope (SOT) on board of the Hinode
space satellite \citep{hinode_overview07,sot_hinode,suematsu_etal_07}. The four 
Stokes profiles of the \ion{Fe}{i}\,630\,nm line pair  were recorded
in a quiet Sun area close to the disc centre. The integration time of 4.8\,s led to 
an rms noise level in the polarization signal of 
$\sigma\approx1.2\times10^{-3}\,I_{\rm{c}}$. 
The data calibration for the SP is described in \cite{ichimoto_etal_07} and \cite{tarbell_06}.
The field of view comprises an area of $162\arcsec\times 302\arcsec$
with a spatial sampling of $0.16\arcsec$ 
along the slit and $0.15\arcsec $ in the scanning direction. 
The spatial resolution of the resulting spectro-polarimetric maps is approximately $0.3\arcsec$,
and the spectral sampling is 2.15\,pm. The rms continuum contrast at 630.0\,nm is around 7.5\,\%, 
far above the 3\,\% value typically observed from the ground at the same 
wavelength \citep{reza_etal_4, khomenko_etal_05L}.
This same data set served \citet{lites_etal_07a,lites_etal_07b} in an investigation of
the horizontal magnetic fields in the quiet Sun. 

\enlargethispage{10pt}

%------------------------------- Fig. 1 ----------------------------------------------
\begin{figure*}
\resizebox{\hsize}{!}{
\includegraphics{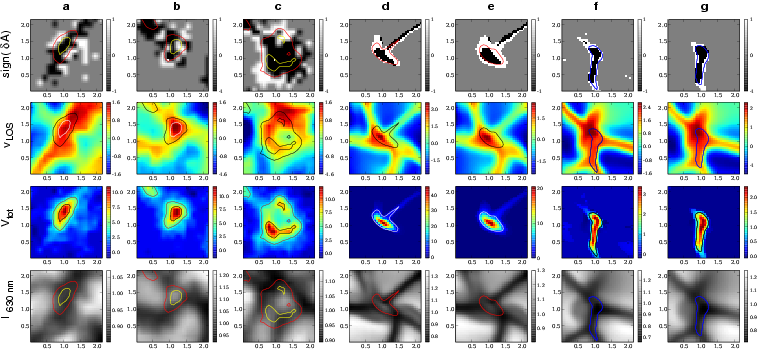}}
\caption{\emph{Columns a-c}: observational data obtained 
   with the spectro-polarimeter of Hinode/SOT. \emph{Columns d and f}: 
   synthetic data from the 3-D MHD simulation. \emph{Columns e and g}:
   same as d and f but after application of the SOT-PSF to the
   synthetic intensity maps. \emph{Rows from top to bottom}: the 
   sign of the Stokes-$V$ area asymmetry for the line \ion{Fe}{i}\,630.25\,nm, where 
   black and white correspond to negative and positive, respectively, the 
   \ion{Fe}{i}\,630.15\,nm line-wing velocity in km\,s$^{-1}$, $V_{\rm{tot}}$
   of \ion{Fe}{i}\,630.25\,nm in m\AA , and the continuum intensity at 630.0\,nm. 
   Contours refer to $V_{\rm{tot}}$. Distance between tick marks 
   is $0.5\arcsec$.}
\label{fig:maps}
\end{figure*}
%-------------------------------------------------------------------------------------

%%%%%%%%%%%%%%%%%%%%%%%%%%% Numerical Simulations %%%%%%%%%%%%%%%%%%%%%%%%%%%%

\section{Numerical simulation} 

The three-dimensional magnetohydrodynamic simulation comprises
an area of $4\,800\times 4\,800$\,km$^\mathrm{2}$, corresponding to
$6.6\arcsec\times 6.6\arcsec$ and spans a height range of 2\,800\,km
from the top layers of the convection zone to the middle chromosphere.
Details of the simulation, carried out with the CO$^\mathsf{5}$BOLD code, 
can be found in \cite{cobold_05, cobold_06}.

Initially, a homogeneous vertical magnetic field with
a strength of 1\,mT was superposed on a model of relaxed 
thermal convection. After relaxation, fields of mixed polarity occur 
throughout the photosphere with an area-fractional imbalance typically of 
3:1 for fields stronger than 1\,mT. Similar polarity imbalances of even
larger fields of view also occur in observational data \citep{lites02}.
Because of the periodic side boundaries and
the conditions that the magnetic field must remain vertical at the 
top and bottom boundaries, the horizontally averaged vertical net 
magnetic-flux density remains 1\,mT throughout the simulation. This value compares
very well with the mean flux density of 1.12\,mT reported
by  \cite{lites_etal_07a} for the Hinode data set. The mean 
absolute horizontal field strength is 2.5\,mT in the upper photosphere
of the simulation, 2.5 times the net vertical flux density.
\citet{lites_etal_07a} report a ratio of 5.

The grid constant of the computational domain in the horizontal direction
is $0.055^{\prime\prime}$ (120$^3$ grid cells), three times 
smaller than the spatial sampling of the Hinode SP data.
The simulation was advanced for 4400\,s of real time. The results 
presented here refer to the last 2400\,s, well after the
initial relaxation phase.
Using the SIR code~\citep{sir92, sir_luis}, we 
synthesized the Stokes profiles of the \ion{Fe}{i}~630\,nm line pair 
from a large number of simulation snapshots  with a spectral 
sampling of 2\,pm. Profiles were computed along vertical 
lines-of-sight for each grid point. To the resulting intensity
maps, we additionally applied a theoretical point-spread function (PSF)
that models the spatial resolution capability of the SOT optics 
\citep{wedemeyer08grancont1}.

%%%%%%%%%%%%%%%%%%%%%%%%%%%%% Data An alysis %%%%%%%%%%%%%%%%%%%%%%%%%%%%%%

\section{Data analysis}

The line parameters extracted from both the calibrated Stokes data of the observation
and from the simulation include the area and amplitude asymmetry of Stokes-$V$ ,
the  line-core and line-wing velocities, and the total (signed) circular polarization
\begin{equation}
  V_{\mathrm{tot}} = \int^{\lambda_0}_{\lambda_{\mathrm b}} 
                        V(\lambda)/I_{\mathrm{c}}\, \mathrm{d}\lambda  
                   - \int^{\lambda_{\mathrm r}}_{\lambda_0} 
                        V(\lambda)/I_{\mathrm{c}}\, \mathrm{d}\lambda\,,
\end{equation} 
where $\lambda_0$ is the zero-crossing wavelength of a regular Stokes-$V$ profile,
and $\lambda_{\mathrm r}$ and $\lambda_{\mathrm b}$ denote  fixed wavelengths in the red and blue 
continua of the lines \citep[similar to][]{reza_etal_4}.
The relative Stokes-$V$ area asymmetry \citep{solanki+stenflo84} is defined as 
\begin{equation}
  \delta A = \left({\int^{\lambda_0}_{\lambda_{\mathrm b}} |V(\lambda)|\,\mathrm{d}\lambda  
            - \int^{\lambda_{\mathrm r}}_{\lambda_0}  |V(\lambda)|\,\mathrm{d}\lambda }\right)\Big/
             {\int^{\lambda_{\mathrm r}}_{\lambda_{\mathrm b}} |V(\lambda)|\,\mathrm{d}\lambda }\,.
\end{equation}

Figure~\ref{fig:maps} shows a few selected examples of magnetic elements from the 
Hinode data (columns a-c) and the simulation (columns d-g). The observed ones are 
all taken from network cell interiors. Columns e and g
are based on the same synthetic data as are columns d and f, respectively, but degraded
with the theoretical PSF for SOT. The 
top row displays the sign of the area asymmetry: black corresponds to negative and white to 
positive area asymmetries. Gray indicates that the Stokes-$V$ profile has a signal 
below the 3$\sigma$ noise level (65 \% of the total field of view) or that it has an
irregular shape (20 \%), where most of the latter would classify as regular when
lowering the noise level.
This same threshold also applies to the simulation data of the panels in columns d-g,
where we ignore $V$-profiles with an amplitude less than $3.6\times 10^{-3}\,I_{\rm{c}}$ 
for comparison with the Hinode data. In the rightmost two panels, this threshold is 5 times 
lower in order to highlight the case of a weak magnetic element that would have barely been
detected with Hinode. 

The line-wing velocity of \ion{Fe}{i}\,630.15\,nm, shown in the second row, 
is obtained by averaging the Stokes-$I$ 
Doppler shift in the intensity interval 0.8-$0.9 I_{\rm c}$. 
We set the convective blue shift of the line-core velocity in the quiet Sun to 
$-0.2$\,km\,sec$^{-1}$. Because of the low degree of polarization, the line-wing 
velocity is not spoilt by magnetic influence.

The third row displays $V_{\mathrm{tot}}$, whose contours are shown in all the panels. 
It is a linear measure of the magnetic flux density according
to \cite{lites_etal_99} and \cite{lites_etal_07b}. From the fourth row, which
shows the continuum intensity at 630\,nm, we see that the magnetic flux concentration occurs
mainly in the intergranular space.

%------------------------------- Fig. 2 ----------------------------------------------
\begin{figure}
\centering
\includegraphics[width=70mm]{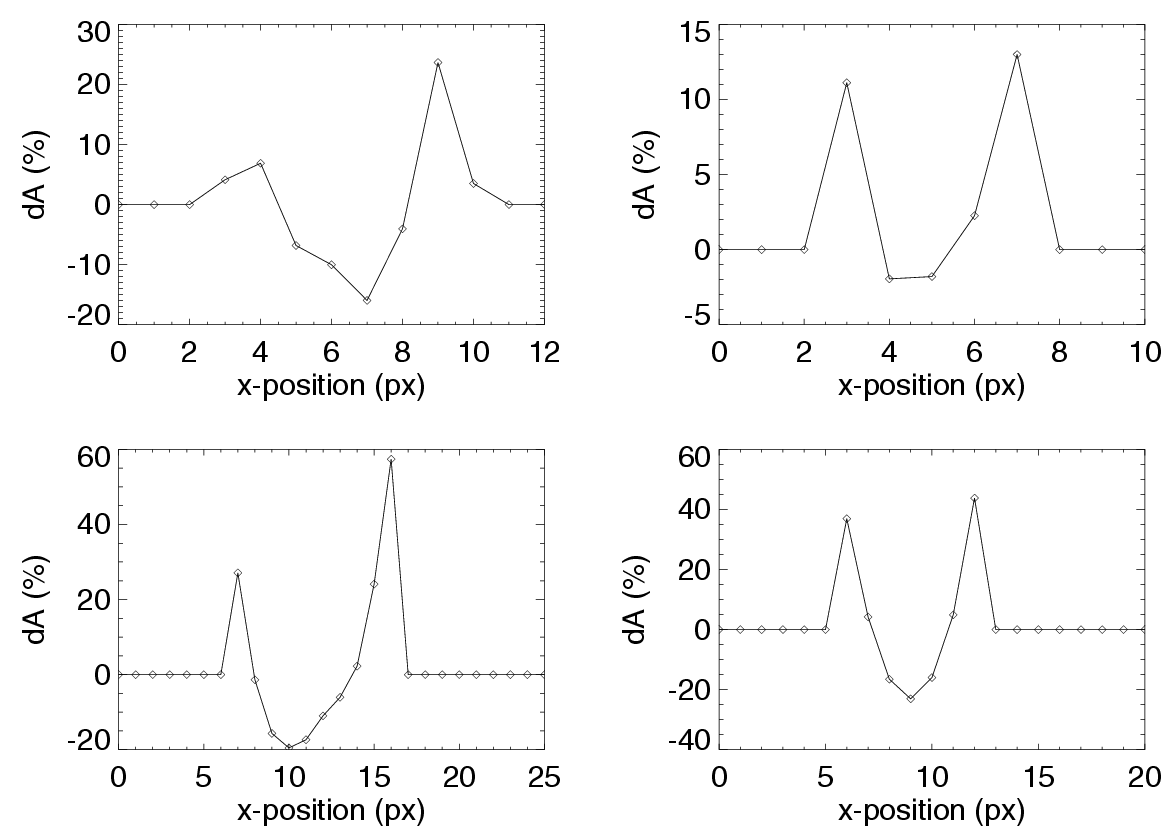}
\caption{Variation in $\delta A$ across magnetic elements from the Hinode
   data \emph{(top row)} and the simulation \emph{(bottom row)}. From left to 
   right and top to bottom, the sections are taken from the data corresponding
   to columns a, b, d, and f of Fig.~\ref{fig:maps}.}
\label{fig:dAofx}
\end{figure}
%-------------------------------------------------------------------------------------

%%%%%%%%%%%%%%%%%%%%%%%%%% Results and Discussions %%%%%%%%%%%%%%%%%%%%%%%%%%%%

\section{Results and discussion} 

\begin{figure*}[t!]
\resizebox{\hsize}{!}{
\includegraphics{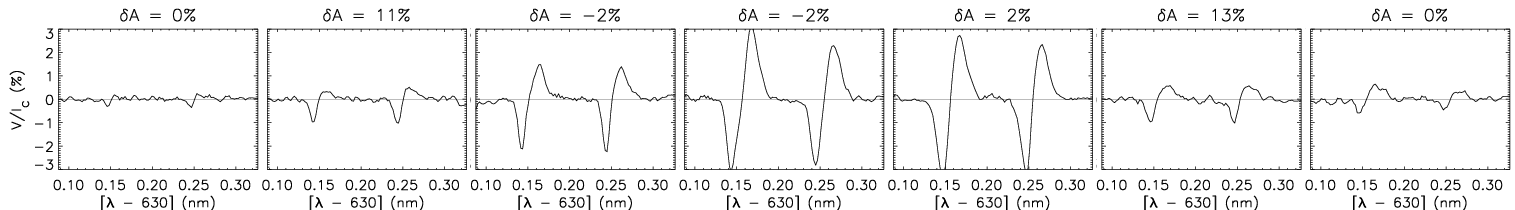}}
\caption{Stokes-$V$ profiles from the Hinode data across the magnetic element
   of column b of Fig.~\ref{fig:maps}. $\delta A(x)$
   for the same section is shown in Fig.~\ref{fig:dAofx} (top right panel).}
\label{fig:profiles_hin}
\end{figure*}

Both the observation and the simulation show magnetic elements of either
sheet like or roundish shape, possessing the same striking pattern with regard
to the Stokes-$V$ asymmetry: a central region with negative Stokes-$V$ area 
asymmetry is framed or surrounded by a narrow seam of pixels having positive area 
asymmetry (Fig.~\ref{fig:maps}, first row). 
Plotting $\delta A$ across a magnetic element, we find that $|\delta A|$ is 
typically larger in the periphery than in the central region of the magnetic flux 
concentration. Representative cross sections that demonstrate this behaviour 
are shown in Fig.~\ref{fig:dAofx} for both the observation and the simulation.
Along with $\delta A$ in the top right panel of Fig.~\ref{fig:dAofx},
Fig.~\ref{fig:profiles_hin} also gives the corresponding observed Stokes-$V$ profiles. 
Also from Fig.~\ref{fig:maps} we see that the magnetic elements are located
within and surrounded by a downdraft. 

%------------------------------- Fig. 3 ----------------------------------------------
\begin{figure}
\centering
\includegraphics[width=73mm]{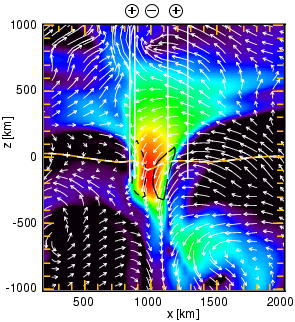}
\caption{Vertical cross section through the simulation box corresponding
   to position $y=1.15\arcsec$ of the column d of Fig.~\ref{fig:maps}.
   It displays the logarithmic magnetic field strength 
   (from 1.0 to $10^{2.3}$~mT, colour-coded), together with the velocity field, 
   projected on the vertical plane (white arrows) and the electric current density
   normal to the plane (black contours). Optical depth $\tau_{500\,{\rm nm}}\! =\! 1$
   is shown as well. The white vertical lines indicate the ranges in which
   Stokes-$V$ profiles from vertical lines-of-sight have either positive or
   negative area asymmetry, $\delta A$. Outside this range, the Stokes-V signal is 
   below the $3\sigma$ noise level of the observations. $\delta A(x)$ is plotted in 
   the lower left panel of Fig.~\ref{fig:dAofx}.}
\label{fig:cut}
\end{figure}
%-------------------------------------------------------------------------------------

For gaining deeper insight, we now consider in Fig.~\ref{fig:cut} the vertical 
cross section through the simulation snapshot shown in column d of 
Fig.~\ref{fig:maps} for the position $y=1.15\arcsec$. It displays the magnetic field 
strength (colour-coded) together with the velocity field, projected on the vertical 
plane (white arrows). The horizontally running curve indicates optical depth  
$\tau_{500\,{\rm nm}}\! =\! 1$.  A magnetic flux concentration 
is forming in the downdraft and has reached a field strength of
0.13~T at $\tau_{500\,{\rm nm}}\! =\! 1$. Close to the surface of optical 
depth unity and through the photosphere, a relatively sharp,  
funnel-shaped boundary separates the magnetic flux concentration from the  
weak-field or field-free surroundings.

The inner vertical white lines indicate a
central part of the flux concentration ($880 < x < 1120$~km), where
Stokes-$V$ profiles have $\delta A < 0$. This is because their lines-of-sight 
sample increasing field strength and increasing downflow velocity with increasing 
continuum optical depth \citep{solanki+pahlke88}. This situation 
changes drastically in the peripheral region $840 \le x < 880$~km
and $1120 < x \le 1280$~km (also indicated with white lines), 
where $\delta A > 0$. Here, the
downflow speed still increases with optical depth; but because of the
funnel-shaped flux concentration, lines-of-sight traverse a magnetic
boundary layer, where the field strength rapidly drops
with increasing optical depth. The combination of increasing velocity
and decreasing field strength produces positive area asymmetry.
From this, we conclude that the narrow seam of positive area asymmetry
in the periphery of magnetic flux elements seen in the Hinode data is
a consequence of their well-defined boundary (or magnetopause)
that expands with height in a funnel-like manner. 

%------------------------------- Fig. 4 ----------------------------------------------
%-------------------------------------------------------------------------------------

As an immediate consequence of Amp\`ere's law, this boundary
carries an electric current sheet. The black contours in
Fig.~\ref{fig:cut} indicate a current density of 0.5\,Am$^{-2}$, 
encircling  higher values of up to 4.0\,Am$^{-2}$. This refers
to the current component perpendicular to the section plane, which
flows in the opposite direction on either side of the
magnetic flux concentration.

If the magnetic flux concentration is inclined enough with respect to the 
viewing direction, lines-of-sight 
traverse the boundary layer on only one side of the flux concentration, which leads
to a one-sided seam of positive area asymmetry. This is almost the case
in Fig.~\ref{fig:cut} (and corresponding $\delta A$-panel of Fig.~\ref{fig:maps}),
where the flux concentration is slightly inclined to the right making the
seam of positive $\delta A$ to the left much narrower than that on the right side.
In fact, most magnetic elements in the field of view of the observation
show a one-sided pattern like in the upper right corner of the synthetic
$\delta A$-map in columns d or e.

Because the mean magnetic flux density in the network cell interior 
is low, granular flow gathers only small amounts of flux so that only small 
and  weak magnetic flux concentrations form \citep{steiner_03}. 
They form and disperse with the onset and  cessation 
of intergranular downdrafts. Therefore we observe strong downflows
associated with the magnetic elements considered in this work. Indeed,
\cite{grossmann_etal_96}, \cite{sigwarth_etal_99}, 
and \cite{martinez-pillet+al97} also 
observe mainly downflows associated with weak magnetic signals.
In combination with increasing field strength, these
downflows give rise to negative area asymmetry. Yet, the peripheral positive 
area asymmetry is larger (Fig.~\ref{fig:dAofx}) and may partially balance
or even outweigh the negative contribution when observing at a lower
spatial resolution.

\cite{langangen_etal_07} finds that the downdraft velocity near the  
edges of magnetic elements is larger than in the central part. Here
we find in both observations and simulation the maximum speed
in the centre of the flux concentration. In the simulation 
this is because the magnetic element forms
as a consequence of an intergranular downdraft, with peak velocity in the 
centre. This is different from the regime of a mature strong flux concentration
for which simulations exhibit veritable downflow jets near their edges but not 
in the centre \citep{steiner_98,sheylag_etal_07}. This regime 
might be more representative of the active region elements observed 
by \cite{langangen_etal_07}.

\citet{grossmann_etal_88,grossmann_etal_89} and \cite{solanki_89} 
first pointed to the 
peripheral, canopy-like magnetopause of magnetic elements as
the origin of the observed positive Stokes-$V$ area asymmetry. 
\cite{steiner99} found this effect at work in a two-dimensional 
MHD-simulation, but also found that the central region of the 
magnetic flux sheet tends to have negative values  and that a delicate 
balance with the peripheral region exists in which the positive values 
may dominate and outweigh the negative ones. These findings have recently 
been confirmed in a three-dimensional environment by \cite{sheylag_etal_07}.
\cite{bellot-rubio+al00} came to a similar conclusion based on
semi-empirical modelling.
\cite{leka+steiner2001} found a similar pattern in the Stokes-$V$ 
area asymmetry in pores and magnetic knots. Here, for the first time, 
we find this pattern in observations of small-scale magnetic elements 
in the network cell interior.

%===========================================================================

\section{Conclusion}

Spectro-polarimetric data of a quiet Sun area close to the disc centre obtained
with SOT on board Hinode were analysed. We find magnetic elements of 
either a roundish (tube) or an elongated (sheet) shape, which show Stokes-$V$ 
profiles of negative area asymmetry in the centre, surrounded or framed by
pixels of larger, positive area asymmetry. 
A comparison with results from 3-D MHD-simulations suggests
that this peculiar pattern in Stokes-$V$ area asymmetry (first predicted
by Steiner 1999) is due to the confined nature of the field in magnetic 
elements with a funnel-shaped boundary layer that gives rise to a steep 
gradient in field strength along the line of sight. 

We also conclude that this kind of magnetic element
of the internetwork is accompanied by electric current sheets.
While these conclusions are evident from the comparison, we cannot 
exclude  the hypothesis that a suitable magnetic structuring on scales not 
resolved by the present observation and simulation would lead to the observed 
pattern in $\delta A$.

\begin{acknowledgements}
Hinode is a Japanese mission developed and launched by ISAS/JAXA, with NAOJ as the domestic 
partner and NASA and STFC (UK) as international partners. It is operated by these agencies 
in co-operation with ESA and NSC (Norway). 
Part of this work was supported by the Deutsche Forschungsgemeinschaft (SCHM 1168/8-1).
\end{acknowledgements} 

\bibliography{da}        

\end{document}